# Tunnelling Spectroscopy of Andreev States in Graphene

Landry Bretheau[1][†]*, Joel I-Jan Wang[1][†]*, Riccardo Pisoni[1], Kenji Watanabe[2], Takashi Taniguchi[2], Pablo Jarillo-Herrero[1]*

[1] Department of Physics, Massachusetts Institute of Technology, 77 Massachusetts Avenue, Cambridge, Massachusetts 02139, United States
[2] National Institute for Materials Science, Namiki 1-1, Tsukuba, Ibaraki 305-0044, Japan
[†] These authors contributed equally to this work.
*e-mail: bretheau@mit.edu; joelwang@mit.edu; pjarillo@mit.edu

**A normal conductor placed in good contact with a superconductor can inherit its remarkable electronic properties[1,2]. This proximity effect microscopically originates from the formation in the conductor of entangled electron-hole states, called Andreev states[3–8]. Spectroscopic studies of Andreev states have been performed in just a handful of systems[9–13]. The unique geometry, electronic structure and high mobility of graphene[14,15] make it a novel platform for studying Andreev physics in two dimensions. Here we use a full van der Waals heterostructure to perform tunnelling spectroscopy measurements of the proximity effect in superconductor-graphene-superconductor junctions. The measured energy spectra, which depend on the phase difference between the superconductors, reveal the presence of a continuum of Andreev bound states. Moreover, our device heterostructure geometry and materials enable us to measure the Andreev spectrum as a function of the graphene Fermi energy, showing a transition between different mesoscopic regimes. Furthermore, by experimentally introducing a novel concept, the supercurrent spectral density, we determine the supercurrent-phase relation in a tunnelling experiment, thus establishing the connection between Andreev physics at finite energy and the Josephson effect. This work opens up new avenues for probing exotic topological phases of matter in hybrid superconducting Dirac materials[16–18].**

When a normal quantum conductor (N) is sandwiched between two superconductors (S), a current can flow in absence of any voltage due to the Josephson effect[1]. This macroscopic supercurrent is driven by the difference between the order parameter phases of the two superconductors $\varphi$. Microscopically, it corresponds to the coherent flow of Cooper pairs through the conductor, which is made possible by successive Andreev reflections at the N-S interfaces, where electrons (holes) are reflected as opposite-spin holes (electrons) (Fig. 1a). Due to this process, resonant electron-hole states form in the central conductor, known as Andreev Bound states (ABS)[3–6]. They lie as energy levels inside the superconducting gap $[-\Delta, \Delta]$, only negative-energy states being populated in the ground state (Fig. 1b). Crucially, the Andreev energy $E_n(\varphi)$ depends on $\varphi$, and each populated ABS carries in response a supercurrent $(1/\phi_0)(\partial E_n/\partial \varphi)$, where $\phi_0 = \hbar/2e$ is the reduced flux quantum. The overall phase-dependent Andreev spectrum is shown in Fig. 1c and depends on geometric and microscopic parameters[7,8], such as the conductor dimensions that determine the number of ABS, the scattering processes in the conductor and the contacts transparency. Strong ABS phase dependence and therefore prominent proximity effect is achieved in ballistic systems with transparent N-S interfaces[7,8].

Graphene (G) can exhibit low contact resistance and weak scattering when connected to superconducting electrodes[19,20]. These properties, combined with the ability to control the number of conduction channels with a gate-voltage, make graphene an ideal test-bed for exploring Andreev physics in 2D. Although the superconducting proximity effect in graphene has attracted considerable research interest[19–29], most of the experimental studies have been limited to transport measurements of dissipationless supercurrent in graphene-based (SGS) Josephson junctions. Accessing the energy domain while controlling the superconducting phase difference is crucial for probing ABS, and it has been performed in just a few systems such as silver wires[9], carbon nanotubes[10], semiconductor nanowires[11], and atomic break-junctions[12,13]. However, a direct spectroscopic observation of phase-dependent ABS in graphene is missing. The way ABS form in graphene still remains an open question, especially around the neutrality point where its unique electronic spectrum should give rise to specular Andreev reflections[27,30] and potentially to exotic ABS. Interest in such a spectroscopy has increased with recent proposals for hosting Majorana modes[16–18] within proximitized graphene operated in the quantum Hall regime[29].

To probe Andreev physics in the energy domain, we perform tunnelling spectroscopy of graphene proximitized by superconductors (Fig. 1d). The experiment is performed using a full Van der Waals heterostructure shown schematically in Figure 1e. A monolayer graphene sheet is encapsulated between hexagonal boron nitride (hBN) crystals. The bottom one is 15 nm thick and isolates graphene from a graphite local back-gate, which enables us to control electrostatically the Fermi energy of graphene with the voltage $V_g$. The top hBN crystal is just one atom thick and is used as a tunnelling barrier. Immediately on top sits a 150 nm-wide metallic probe made of thin graphite. The use of a graphite probe is crucial, as it limits the doping in the graphene (directly underneath the monolayer hBN barrier) owing to the small work function difference between graphite and graphene. This makes the low-density regime around the charge neutrality point (CNP) accessible to our investigations. The graphene sheet, with width $W = 2\ \mu m$, is well connected to two superconducting aluminium electrodes, with an inter-lead distance $L = 380\ nm$. These electrodes are patterned in a loop that enables us to control the phase difference $\varphi = \Phi/\phi_0$ across the graphene by applying a magnetic flux $\Phi$ through the loop.

By measuring the differential conductance $dI/dV$ of this tunnelling device at low temperature (20 mK), one can infer the local density of states (DOS) of graphene at energy $eV$, $V$ being the bias voltage between the graphite probe and the superconducting electrodes (Fig. 1d-e). Such a measurement is shown in Fig. 2a for different values of the magnetic field and a constant gate voltage $V_g = 1\ V$. Each trace exhibits a dip centred at zero bias and two side peaks at $V = \pm 160\ \mu V$. We emphasize that the tunnelling process takes place between thin graphite (i.e. non-superconductor) and the graphene flake, which is only laterally contacted by the superconductors. The graphene DOS, usually featureless within this very narrow energy range in the normal state, has dramatically changed due to superconducting proximity effect and displays a soft induced gap $\Delta \sim 160\ \mu eV$, only slightly reduced compared to the one of aluminium $\Delta_{Al} \sim 180\ \mu eV$. When varying the magnetic field $B$ (Fig. 2b), the DOS oscillates with a periodicity $\delta B = 360\ \mu T$, which corresponds to one flux quantum $2\pi\phi_0$ threading the loop of enclosed area $A = 5.7\ \mu m^2$ (taking into account Meissner effect). The corresponding phase $\varphi = A(B - B_0)/\phi_0$ is shown on the top axis, where $B_0 = 250\ \mu T$ is an offset magnetic field in this experiment.

The gap and side peaks are most pronounced at $\varphi = 0$ and get reduced when the phase is swept towards $\varphi = \pi$ (see Fig. 2b, top axis).

This flux-dependent DOS is consistent with a continuum of ABS modulating with phase. The graphene weak link being spatially extended indeed consists of a large number of conduction channels $M \sim 2W/\lambda_F$, each containing $\sim 2L_m/\pi\xi$ pairs of ABS[8]. Here, $\lambda_F = 2\sqrt{\pi e/c_g|V_g|}$ is the Fermi wavelength of electrons in graphene (with $c_g \approx 2.3\ fF/\mu m^2$ the gate capacitance per unit area) and $M$ can be tuned with the back-gate $V_g$ typically between 40 and 300, $\xi$ is the superconducting coherence length and $L_m$ the effective length of channel $m$. To estimate the superconducting coherence length $\xi$, we measured in the normal regime the gate-dependent differential resistance of a similarly fabricated graphene junction (see Supplementary Fig.S5). From the extracted mean free path $l_e \sim 140\ nm < L$, one can infer a lower-bound $\xi \sim 590\ nm$ using the diffusive relation $\xi = \sqrt{\hbar D/\Delta}$, with $D = v_F l_e/2$ the Einstein diffusion coefficient in 2D and $v_F$ the Fermi velocity. Since the S-G-S junction is in the intermediate regime $L \sim \xi$, the spectrum is complex and made out of different type of ABS. Most channels are in the short junction regime ($L_m < \pi\xi/2$) and contain a single pair of ABS (solid lines in Fig. 1c). In the case of good coupling to the superconducting leads, the ABS strongly depend on phase and are responsible for the flux modulation of the DOS. The latter indeed peaks strongly at the gap when $\varphi = 0$ due to the accumulation of ABS at energies $E_n \sim \pm\Delta$. At $\varphi = \pi$, the ABS reach their minimum energy and populate the gap, resulting in an enhanced DOS at low energy. At the same time, channels with low contact transparency provide ABS that detach from the gap edges at $\varphi = 0$ and have weaker phase modulation, thus explaining the observed soft superconducting gap. On top of that, channels with large transverse momentum are effectively longer and contain more ABS that fill the superconducting gap and which exhibit weaker phase modulation (dashed lines in Fig. 1c), contributing therefore to the finite DOS in the soft gap region.

To characterize further how Andreev physics manifests in graphene, we now vary the carrier density of the weak link $n = c_g|V_g|/e$. In the normal state, the differential conductance as a function of $V_g$ is V-shaped with a minimum at $V_{CNP} \sim 0.05\ V$, which reflects the linear DOS of the Dirac cone (see Supplementary Fig. S4). Fig. 3 shows the DOS in the superconducting state, as a function of both energy and phase, for various values of $V_g$ (see also Supplementary Fig. S3 for a more exhaustive gallery). The energy spectra depend strongly on the graphene carrier density, which is tuned from hole-type (1st row) through the CNP (2nd row) to electron-type (3rd row). Noticeably, the larger the carrier density, the stronger the DOS is modulated with phase, resulting in a larger supercurrent which is in agreement with transport measurements of G-based Josephson junctions[19–23]. In contrast, at low density $|n| < 1 \times 10^{11}/cm^2$, close to the CNP (Fig. 3e), the phase modulation of the DOS is very weak. In this gate voltage range, which corresponds to $|E_F| < 37\ meV$ in Fermi energy, the normal state DOS also shows negligible $V_g$ dependence. We attribute these effects to disorder, which causes the formation of electron-hole puddles in graphene when its energy scale is larger than the Fermi energy[31,32]. In such a heterogeneous and disordered configuration[25], the mean free path is reduced and the graphene enters a long junction regime. Moreover, the coupling of a given puddle to the superconducting leads tends to be reduced and asymmetric. This results in an overall weak phase modulation of the DOS. On the contrary, at large carrier density the phase modulation of the DOS is much stronger. This is consistent with a disorder potential smaller than the Fermi energy, which results in more ABS from well-coupled and transmitted conduction channels in the short junction limit. At very large carrier density $|n| > 3{\times}10^{12}/cm^2$, one can even measure a complete closing of the induced gap and a flat DOS at $\varphi = \pi$ (Fig. 3a,h,i, and Supplementary Fig. S5), as more ballistic ABS reach zero energy at $\varphi = \pi$ (solid lines in Fig. 1c).

The effect of the normal scattering properties also appears as an asymmetry between the energy spectra for opposite carrier density (see Fig. 3 and Supplementary Fig. S6). When the graphene is hole-doped, the phase modulation of the DOS is indeed smaller with a V-shaped induced gap, whereas it is U-shaped with a stronger phase modulation for the electron-doped case. This is because aluminium n-dopes graphene underneath the contact due to their work function difference, which results in an n-p-n potential profile when $V_g < V_{CNP}$. These p-n junctions reduce the contacts' transparency, which repel the ABS from the gap edges toward lower energy (solid lines in Fig. 1c) and weaken the phase modulation of the DOS, in good agreement with measurements of Al-G-Al Josephson junctions that shows smaller supercurrent in the hole-doped region[21].

Probing Andreev physics in the energy domain while controlling the phase difference additionally allows one to access the supercurrent spectral density $J_S(E, \varphi)$, which quantifies the amount of supercurrent carried by Andreev states at energy $E$ and phase $\varphi$. This quantity makes evident the link between Andreev physics and Josephson effect. It has seldom been discussed in the literature[33–36] and, to the best of our knowledge, has never been directly connected to the DOS. We propose here the heuristic definition $J_S(E, \varphi) = -\frac{1}{\phi_0} \int_0^E \frac{\partial DOS}{\partial \varphi}(\epsilon, \varphi) d\epsilon$, which can easily be checked as being correct in the case of a set of discrete ABS. Using this definition and the factor $c = 34.7\ meV^{-1}\mu S^{-1}$ to convert tunnelling conductance to DOS (see section 6 in Supplementary Information), one can numerically compute $J_S$ for each energy spectrum. Fig. 4 shows both the Andreev spectrum in graphene and its corresponding supercurrent spectral density for a gate voltage $V_g = 2.4\ V$. Remarkably, the data confirm that opposite energy ABS carry opposite supercurrent. Going further, one can derive the supercurrent in the ground state defined as $I_S(\varphi) = \frac{1}{2}\int_{-\infty}^{\infty} \text{sgn}(-E) J_S(E, \varphi)\, dE$ (Fig. 4c) and thus access the Josephson current-phase relation. In this measurement, which was taken over a large range in magnetic field, both DOS, $J_S$ and $I_S$ oscillate with phase, the oscillations being progressively washed out at large field. At large fields, a significant flux is threading graphene due to its extended 2D-nature, which results in a space-dependent phase difference and a dephasing of ABS along the junction. Close to zero flux, the current-phase relation is strongly anharmonic and can be fitted by the supercurrent carried by perfectly coupled short ABS with an average transmission coefficient $\tau = 0.725$ (solid red curve)[6]. This confirms that ABS from well-coupled and transmitted short channels are dominating the supercurrent at large carrier density.

Our tunnelling spectroscopy study provides fundamental insights into how Josephson effect develops in graphene, and it can be extended to other 2D materials. The combined effects of disorder, geometry, and Fermi energy offer many opportunities to investigate Andreev physics in different mesoscopic superconductivity regimes. Moreover, as graphene's extended two-dimensional nature enables one to combine superconductivity and the quantum Hall effect[29], this platform is promising for the detection of Majorana modes[18], key ingredients for topologically protected quantum computation.

## Methods
The van der Waals heterostructure is created via successive pick-ups and transfers using a polymer-based dry transfer technique[37,38] (see Supplementary Information for a detailed description of the fabrication procedure). The superconducting electrodes are fabricated by electron-beam lithography and evaporation of 70 nm of aluminium, with 7 nm of titanium as a sticking layer. The sample is anchored to the mixing chamber of a dilution refrigerator at 20 mK, well below the transition temperature of aluminium ($T_C \sim 1.1$ K). The biasing and measurement lines are heavily filtered using both discrete and distributed cryogenic low-pass filters[39]. The measurements are made at low frequency (10-100 Hz) using room-temperature amplification and standard lock-in techniques with an excitation voltage of 5-10 µV.

## Acknowledgements
We acknowledge helpful discussions with W. Belzig, J.C. Cuevas, V. Fatemi, Ç. Girit, P. Joyez, A. Levy Yeyati, J-D. Pillet, H. Pothier, V. Shumeiko, and C. Urbina. This work has been primarily supported by the US DOE, BES Office, Division of Materials Sciences and Engineering under Award DE-SC0001819 and by the Gordon and Betty Moore Foundation's EPiQS Initiative through Grant GBMF4541 to P.J.H. J.I-J.W. was partially supported by a Taiwan Merit Scholarship TMS-094-1-A-001. This work made use of the MRSEC Shared Experimental Facilities supported by NSF under award No. DMR-0819762 and of Harvard's CNS, supported by NSF under Grant ECS-0335765.

## Author contributions
J.I-J.W, L.B. and P.J.H. designed the experiment. J.I.-J.W and R.P. fabricated the devices. L.B. and J.I-J.W carried out the measurements. L.B. analysed and interpreted the data. K.W. and T.T. supplied hBN crystals. L.B. and J.I.-J.W wrote the manuscript with input from all the authors.

## Additional information
Supplementary information is available in the online version of the paper. Reprints and permissions information is available online at www.nature.com/reprints. Correspondence and requests for materials should be addressed to L.B., J.I-J.W, and PJH.

## Competing financial interests
The authors declare no competing financial interests

1. Josephson, B. D. Possible new effects in superconductive tunnelling. *Phys. Lett.* **1,** 251–253 (1962).
2. De Gennes, P. G. Boundary effects in superconductors. *Rev. Mod. Phys.* **36,** 225–237 (1964).
3. Kulik, I. O. Macroscopic quantization and the proximity effect in S-N-S junctions. *Sov. Phys. JETP* **30,** 944–950 (1970).
4. Furusaki, A. & Tsukada, M. Dc Josephson effect and Andreev reflection. *Solid State Commun.* **78,** 299–302 (1991).
5. Beenakker, C. W. J. & van Houten, H. Josephson current through a superconducting quantum point contact shorter than the coherence length. *Phys. Rev. Lett.* **66,** 3056–3059 (1991).
6. Bagwell, P. F. Suppression of the Josephson current through a narrow, mesoscopic, semiconductor channel by a single impurity. *Phys. Rev. B* **46,** 12573–12586 (1992).
7. Wendin, G. & Shumeiko, V. S. Josephson transport in complex mesoscopic structures. *Superlatt. and Microstruct.* **20,** 569–573 (1996).
8. Samuelsson, P., Lantz, J., Shumeiko, V. S. & Wendin, G. Nonequilibrium Josephson current in ballistic multiterminal SNS junctions. *Phys. Rev. B* **62,** 1319–1337 (2000).
9. Le Sueur, H., Joyez, P., Pothier, H., Urbina, C. & Esteve, D. Phase controlled superconducting proximity effect probed by tunneling spectroscopy. *Phys. Rev. Lett.* **100,** 197002 (2008).
10. Pillet, J.-D., Quay, C., Morfin, P., Bena, C. & Levy Yeyati, AJoyez, P. Revealing the electronic structure of a carbon nanotube carrying a supercurrent. *Nat. Phys.* **6,** 965–969 (2010).
11. Chang, W., Manucharyan, V. E., Jespersen, T. S., Nygård, J. & Marcus, C. M. Tunneling spectroscopy of quasiparticle bound states in a spinful josephson junction. *Phys. Rev. Lett.* **110,** 217005 (2013).
12. Bretheau, L., Girit, Ç. Ö., Pothier, H., Esteve, D. & Urbina, C. Exciting Andreev pairs in a superconducting atomic contact. *Nature* **499,** 312–315 (2013).
13. Bretheau, L., Girit, Ç. Ö., Urbina, C., Esteve, D. & Pothier, H. Supercurrent Spectroscopy of Andreev States. *Phys. Rev. X* **3,** 041034 (2013).
14. Castro Neto, A. H., Guinea, F., Peres, N. M. R., Novoselov, K. S. & Geim, A. K. The electronic properties of graphene. *Rev. Mod. Phys.* **81,** 109–162 (2009).
15. Das Sarma, S., Adam, S., Hwang, E. H. & Rossi, E. Electronic transport in two-dimensional graphene. *Rev. Mod. Phys.* **83,** 407–470 (2011).
16. Lindner, N. H., Berg, E., Refael, G. & Stern, A. Fractionalizing majorana fermions: Non-abelian statistics on the edges of abelian quantum hall states. *Phys. Rev. X* **2,** 041002 (2012).
17. Clarke, D. J., Alicea, J. & Shtengel, K. Exotic non-abelian anyons from conventional fractional quantum Hall states. *Nat. Commun.* **4,** 1348 (2013).
18. San-Jose, P., Aguado, R., Guinea, F., Fernandez-Rossier, J. & Lado, J. Majorana Zero Modes in Graphene. *Phys. Rev. X* **5,** 041042 (2015).
19. Calado, V. E. *et al.* Ballistic Josephson junctions in edge-contacted graphene. *Nat. Nanotechnol.* **10,** 761–764 (2015).
20. Ben Shalom, M. *et al.* Quantum oscillations of the critical current and high-field superconducting proximity in ballistic graphene. *Nat. Phys.* **12,** 318–322 (2015).
21. Heersche, H. B., Jarillo-Herrero, P., Oostinga, J. B., Vandersypen, L. M. K. & Morpurgo, A. F. Bipolar supercurrent in graphene. *Nature* **446,** 56–59 (2007).
22. Du, X., Skachko, I. & Andrei, E. Y. Josephson current and multiple Andreev reflections in graphene SNS junctions. *Phys. Rev. B* **77,** 184507 (2008).
23. Girit, Ç. *et al.* Tunable graphene dc superconducting quantum interference device. *Nano Lett.* **9,** 198–199 (2009).

24. Dirks, T. *et al.* Transport Through Andreev Bound States in a Graphene Quantum Dot. *Nat. Phys.* **7,** 386–390 (2011).
25. Komatsu, K., Li, C., Autier-Laurent, S., Bouchiat, H. & Guéron, S. Superconducting proximity effect in long superconductor/graphene/ superconductor junctions: From specular Andreev reflection at zero field to the quantum Hall regime. *Phys. Rev. B* **86,** 115412 (2012).
26. Allen, M. T. *et al.* Spatially resolved edge currents and guided-wave electronic states in graphene. *Nat. Phys.* **12,** 128–133 (2015).
27. Efetov, D. K. *et al.* Specular interband Andreev reflections at van der Waals interfaces between graphene and NbSe2. *Nat. Phys.* **12,** 328–332 (2015).
28. Natterer, F. D. *et al.* Scanning tunneling spectroscopy of proximity superconductivity in epitaxial multilayer graphene. *Phys. Rev. B* **93,** 045406 (2016).
29. Amet, F. *et al.* Supercurrent in the quantum Hall regime. *Science* **352,** 966–969 (2016).
30. Beenakker, C. Specular Andreev Reflection in Graphene. *Phys. Rev. Lett.* **97,** 067007 (2006).
31. Martin, J. *et al.* Observation of electron–hole puddles in graphene using a scanning single-electron transistor. *Nat. Phys.* **4,** 144–148 (2008).
32. Xue, J. M. *et al.* Scanning tunnelling microscopy and spectroscopy of ultra-flat graphene on hexagonal boron nitride. *Nat. Mater.* **10,** 282–285 (2011).
33. van Wees, B. J., Lenssen, K.-M. H. & Harmans, C. J. P. M. Transmission formalism for supercurrent flow in multiprobe superconductor-semiconductor-superconductor devices. *Phys. Rev. B* **44,** 470–473 (1991).
34. Wilhelm, F. K., Schön, G. & Zaikin, A. D. Mesoscopic Superconducting-Normal Metal-Superconducting Transistor. *Phys. Rev. Lett.* **81,** 1682–1685 (1998).
35. Baselmans, J. J. A., Morpurgo, A. F., van Wees, B. J. & Klapwijk, T. M. Reversing the direction of the supercurrent in a controllable Josephson junction. *Nature* **397,** 43–45 (1999).
36. Baselmans, J. J. A., Heikkilä, T. T., van Wees, B. J. & Klapwijk, T. M. Direct observation of the transition from the conventional superconducting state to the pi state in a controllable Josephson junction. *Phys. Rev. Lett.* **89,** 207002 (2002).
37. Wang, L. *et al.* One-dimensional electrical contact to a two-dimensional material. *Science* **342,** 614–7 (2013).
38. Wang, J. I. *et al.* Electronic Transport of Encapsulated Graphene and WSe 2 Devices Fabricated by Pick-up of Prepatterned hBN. *Nano Lett.* **15,** 1898–1903 (2015).
39. Spietz, L., Teufel, J. & Schoelkopf, R. J. A Twisted Pair Cryogenic Filter. *arXiv* 1–12 (2006). Preprint at http://arxiv.org/abs/cond-mat/0601316.
40. Wang, L. *et al.* One-dimensional electrical contact to a two-dimensional material. *Science* **342,** 614–617 (2013).
41. Zomer, P. J., Guimarães, M. H. D., Brant, J. C., Tombros, N. & van Wees, B. J. Fast pick up technique for high quality heterostructures of bilayer graphene and hexagonal boron nitride. *Appl. Phys. Lett.* **105,** 013101 (2014).
42. Gorbachev, R. V *et al.* Hunting for monolayer boron nitride: optical and Raman signatures. *Small* **7,** 465–468 (2011).
43. Amet, F. *et al.* Tunneling spectroscopy of graphene-boron-nitride heterostructures. *Phys. Rev. B* **85,** 073405 (2012).

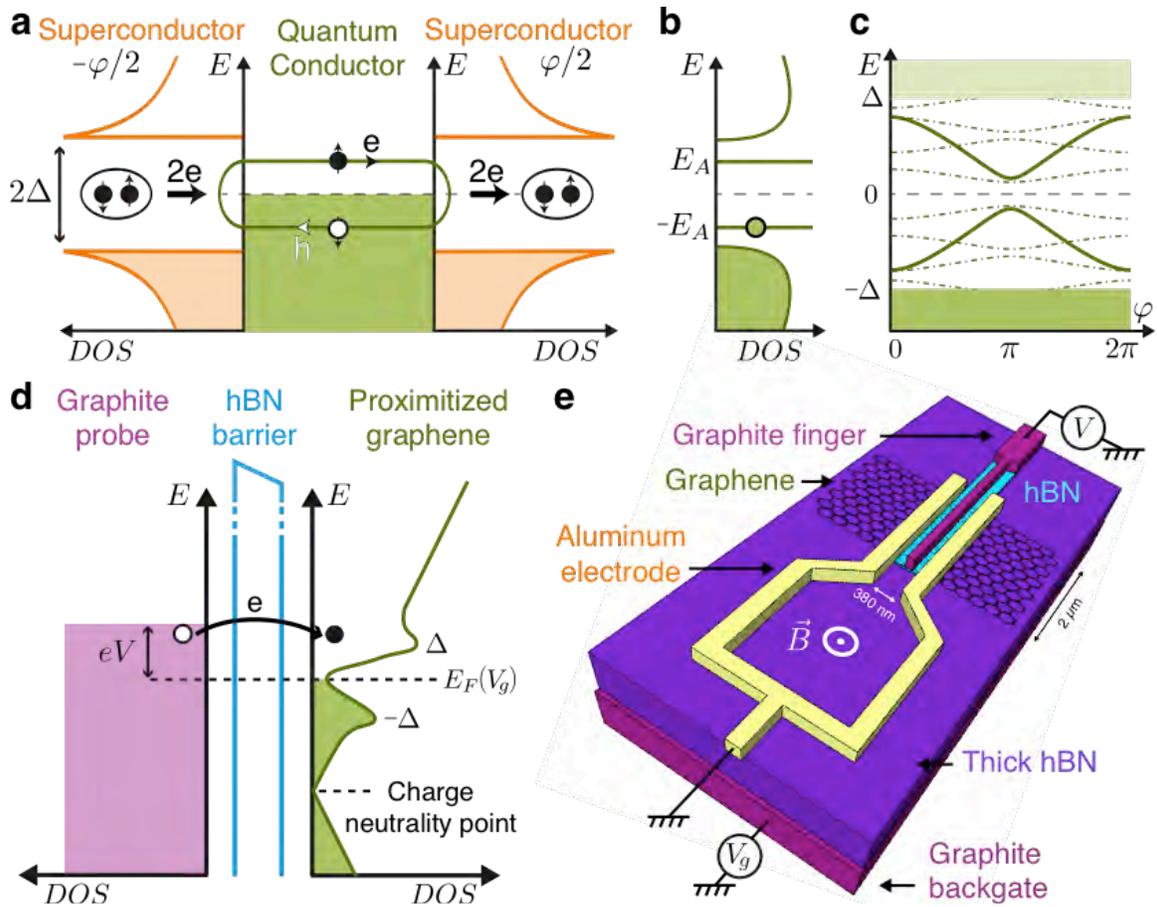

**Figure 1. Experimental concept and device schematic. a**, Microscopic picture of the Josephson effect. Supercurrent – coherent flow of Cooper pairs - through a quantum conductor results from successive Andreev reflections at the normal-superconductor interfaces. **b**, The DOS in the proximitized conductor is dramatically changed with the appearance of resonant ABS within the induced superconducting gap $[-\Delta, \Delta]$. **c**, ABS' energies as a function of phase, for different configurations of the central conductor. In the case of a short single-channel (solid line), there is a single pair of ABS, whereas a long junction contains a large number of ABS (dashed lines). Finite transmission $\tau_S < 1$ in the conductor leads to a repulsion of the ABS away from zero energy at $\varphi = \pi$, and a reduced contact transparency $\tau_C < 1$ detaches the ABS from the gap edges at $\varphi = 0$. **d**, Schematics of the tunnelling spectroscopy process. The normal probe is a graphite electrode and the tunnelling barrier a monolayer hBN crystal. **e**, Device schematics. An encapsulated graphene flake is connected to two superconducting electrodes. Magnetic flux $\phi$ threading the loop imposes a phase $\varphi = \phi/\phi_0$ across graphene and modulates the Andreev states energy.

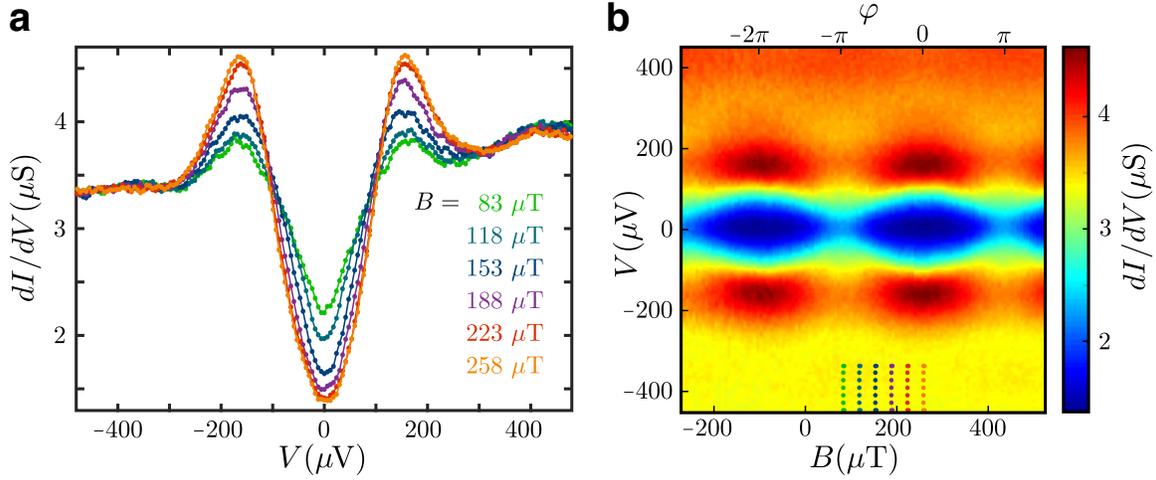

**Figure 2. Phase dependence of the graphene DOS. a**, Differential conductance, *dI/dV* as a function of bias voltage, *V*, for different magnetic fields, *B*, and at a gate voltage $V_g = 1\,V$. **b**, Colour-coded *dI/dV* as a function of *V* and *B* at $V_g = 1\,V$ (the top axis shows *B* converted into the superconducting phase difference $\varphi$). The oscillating spectrum is evidence for a continuum of ABS.

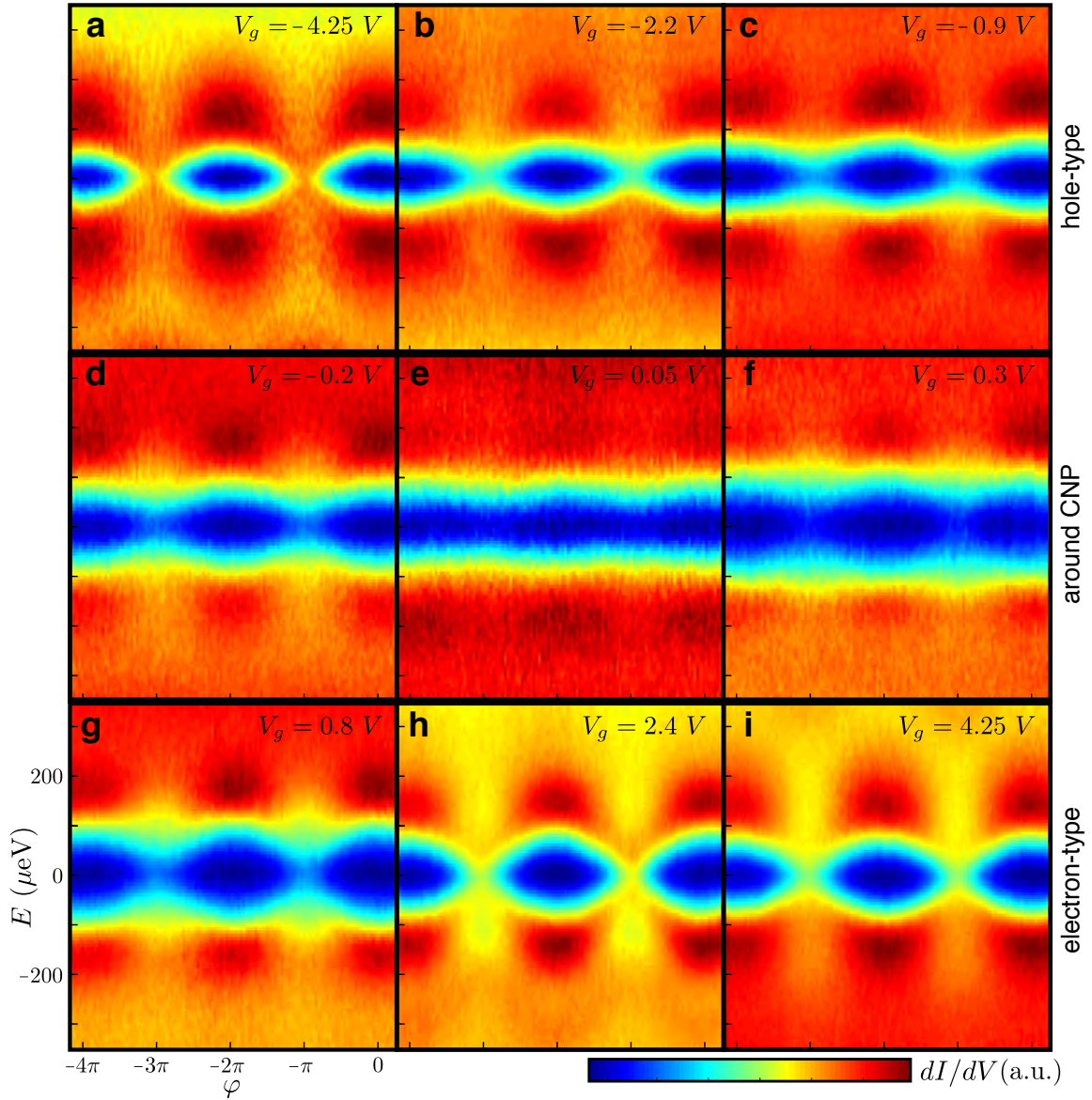

**Figure 3. Gate dependence of the graphene proximitized DOS. a-i**, Normalized differential conductance as a function of both energy $E=eV$ and superconducting phase difference $\varphi$, for different gate voltages, $V_g$ (indicated in each panel). In each panel, the colour-coded differential conductance is linearly scaled to maximize the contrast (see Fig.S4 for quantitative values).

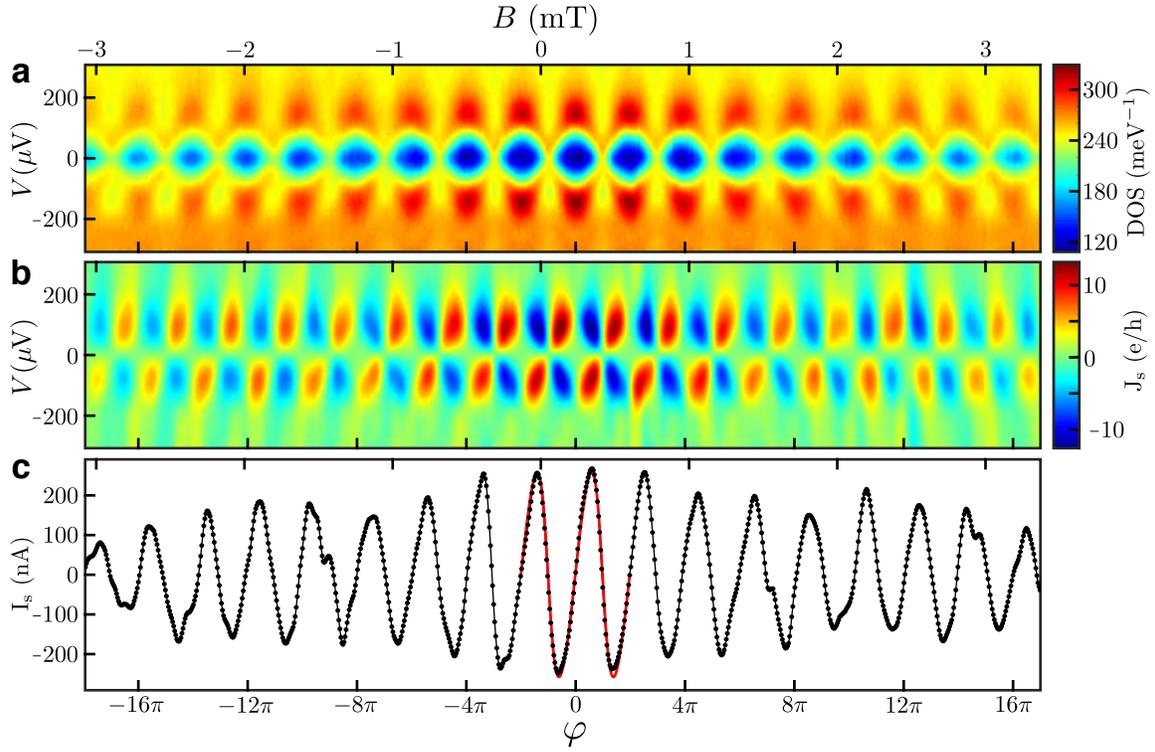

**Figure 4. Andreev states, supercurrent spectral density, and supercurrent in graphene.** Colour-coded DOS (**a**) and supercurrent spectral density $J_S$ (**b**) as a function of both energy $E=eV$ and superconducting phase difference $\varphi$, at a gate voltage $V_g = 2.4\,V$. **c,** Corresponding ground state supercurrent (black dotted line) as a function of $\varphi$. A clear anharmonic supercurrent-phase relation can be seen near zero $\varphi$. The solid red line corresponds to the supercurrent carried by perfectly coupled short ABS with an average transmission coefficient $\tau = 0.725$.

# Supplementary Information for
# "Tunnelling Spectroscopy of Andreev States in Graphene"

Landry Bretheau[1†]*, Joel I-Jan Wang[1†], Riccardo Pisoni[1,2], Kenji Watanabe[3], Takashi Taniguchi[3], Pablo Jarillo-Herrero[1]*

[1] Department of Physics, Massachusetts Institute of Technology, 77 Massachusetts Avenue, Cambridge, Massachusetts 02139, United States
[2] Politecnico di Milano, Piazza Leonardo da Vinci, 32, 20133 Milano, Italy
[3] National Institute for Materials Science, Namiki 1-1, Tsukuba, Ibaraki 305-0044, Japan
[†] These authors contributed equally to this work.
*e-mail: bretheau@mit.edu; joelwang@mit.edu; pjarillo@mit.edu

**Content**

1. Device Fabrication
2. Measurement Setup
3. Scattering Characterization via Transport Measurements
4. Graphene's Density of States in the Normal Regime
5. Graphene's Proximitized DOS in the Superconducting Regime
6. Derivation of Supercurrent Spectral Density

## 1. Device Fabrication

The experiment was performed using a van der Waals heterostructure schematized in Fig.S1a. From top to bottom, it consists of a thin and narrow graphite tunnelling probe (width = 150 nm), an hBN monolayer, a graphene monolayer, an hBN bottom layer (thickness =15 nm), and a graphite bottom gate. To assemble this stack with various thin films, we apply a polymer-based dry pick-up and transfer approach[38,40] using a polycarbonate (PC) film[41] (Fluka Analytical, part # 181641) to achieve higher efficiency in picking up monolayer graphene or hBN from the $SiO_2$ substrate.

To start with, pre-patterned graphite probes on $SiO_2$ substrate are picked-up and transferred onto a hBN monolayer previously identified with Raman spectroscopy[42]. The stack is then shaped into isolated rectangles, each of which defines the length $L$ of a graphene weak link. To pattern the thin films, we perform electron-beam lithography using Poly(methyl methacrylate) (PMMA 950A5 from Microchem) as both the resist and etching mask (thickness ~ 250 nm). After lithography, the graphite is etched by reactive ion etching in oxygen (15 sccm), whereas the hBN is etched in an oxygen (5 sccm), argon (10 sccm), and $CHF_3$ (10 sccm) environment.

We then employ another PC film to successively pick up the probe/tunnelling barrier stack, a monolayer graphene flake, and an hBN bottom layer. All of this is eventually transferred onto a graphite flake deposited on the $SiO_2$ substrate (285 nm thermal oxide on p-doped Si substrate from NOVA electronics). After dissolving the PC film in chloroform, the heterostructure is annealed in forming gas (Ar/$H_2$) at 350 °C for at least 3 hours in order to remove organic residue and to reduce the area with bubbles. Fig.S1b shows an AFM micrograph of the assembled stack.

At last, we perform electron-beam lithography and we deposit 7 nm of Ti and 70 nm of Al by thermal evaporation. After lift-off in acetone, the device is ready for measurement (see Fig.S1c).

## 2. Measurement Setup

Figure S2 shows the schematics of our measurement setup. The sample is anchored to the mixing chamber of a dilution refrigerator at 20 mK, well below the transition temperature of aluminium ($T_C$ ~ 1.1K). All DC lines are heavily filtered using both discrete RC filters and distributed low-pass copper tape filters[39], which are placed at the cryogenic level before reaching the device. $dI/dV$ measurements are performed at low frequency (10-100 Hz) using room-temperature amplification and standard lock-in techniques with an excitation voltage of 5-10 µV.

## 3. Scattering Characterization via Transport Measurements

To estimate the scattering properties of graphene in the normal regime, it's useful to measure in transport a graphene-based junction. However, since the superconducting electrode shunts the graphene flake within a SQUID loop, one cannot directly measure in transport the graphene probed by tunneling spectroscopy. Instead, we measure a graphene junction between two neighboring loops on a device that is fabricated using the same procedure as the device presented in the main text.

Figure S3a shows the normal state resistance $R_N$ of this graphene junction ($L \times W = 350\ nm \times 1\ \mu m$) as a function of backgate voltage. A lower bound of the mean free path, $l_e$, can be estimated using the formula[22] $l_e = \sigma h/(2e^2 k_F)$, where $h$ is Planck's constant, $e$ the electron charge and $\sigma = L/(W R_N)$ the conductivity extracted from the measurement. The Fermi wave vector $k_F = \sqrt{\pi c_g |V_g|/e}$ depends on the backgate voltage $V_g$, with $c_g \approx 2.3 \, fF/\mu m^2$ the gate capacitance per unit square. The extracted mean free path is plotted in Fig.S3b. In the n-doped region ($V_g > 0$), one gets $l_e \sim 140 \, nm \sim L/2.7$, which suggests that the junction is neither ballistic nor diffusive but in an intermediate regime. Note that the extracted $l_e$ is dramatically suppressed in the p-doped region. This is due to p-n junctions that form at the metal/graphene interface and does not reflect the intrinsic quality of the graphene. To infer the superconducting length ξ, we use the formula for the (worst case) diffusive relation $\xi = \sqrt{\hbar D/\Delta}$, with $D = v_F l_e/2$ the Einstein diffusion coefficient in 2D. We find ξ~590 nm, i.e. $L/\xi$~0.65.

Note that in contrast to the graphene flake within the loop, which is encapsulated by a monolayer hBN and the graphite probe, the graphene between two loops is not covered with any other material. This means that this junction is more likely to be contaminated by residue from nanofabrication, and the extracted transport properties such as mean free path or coherence length should be regarded as a lower bound for the actual weak link where tunneling spectroscopy is performed. Based on this estimate, the graphene-based Josephson junction should rather be in the short junction limit (but not infinitely short), and each conduction channel should contain 1 or 2 pairs of ABS[7,8].

## 4. Graphene's Density of States in the Normal Regime

Graphene's normal state DOS is shown as a function of the backgate voltage in Fig. S4a. The overall V-shape curve reflects the DOS corresponding to the linear Dirac cone dispersion, with a minimum at the CNP $V_{CNP} \sim 0.05 \, V$, thus demonstrating the weak doping induced by the graphite probe. On top of it, one can see sharp resonances, which disperse as a function of both energy and backgate voltage in the manner of Coulomb diamonds[43] (see Fig.S4b). They are probably associated with spurious quantum dots at the interfaces with hBN. A detailed analysis suggests that they correspond to 10-20 nm size features with typical addition energy of ~5-15 meV. This energy scale is huge compared to the superconducting gap that appears as a narrow dip at zero energy, and these quantum dots features have little interplay with the superconducting proximity effect and do not affect our measurements.

## 5. Graphene's Proximitized DOS in the Superconducting Regime

We now switch to additional measurements in the superconducting regime.

- Figure S5 shows the DOS in the superconducting state, as a function of both energy and phase, for various values of the gate voltage. The energy spectra depend strongly on the graphene carrier density, which is tuned from hole-type through the CNP to electron-type. As already mentioned in the main text, the larger the carrier density, the stronger the DOS modulates with phase. On top of that, some spectra (a, d, w, x) are dramatically different. There, the differential conductance is modified due to neighbouring quantum dot resonances that create voltage-dependent backgrounds.

- To get rid of this effect, we subtract a bias voltage dependent background defined as the average DOS over one period in magnetic field. The corresponding subtracted spectra are shown in Fig. S6. After subtraction, all spectra share similar features as they display a checkerboard pattern that weakly depends on the carrier density.

- At very large carrier density $|n| > 3 \times 10^{12}/cm^2$, one can measure a complete closing of the induced gap and a flat DOS at $\varphi = \pi$ (Fig. S4 b,u,v). This is emphasized in Fig. S7, which shows the DOS at a gate voltage $V_g = 2.4\,V$. There, disorder is negligible and more ballistic ABS reach zero energy at $\varphi = \pi$.

- Fig. S8 compares energy spectra of opposite carrier density. When the graphene is hole-doped, the phase modulation of the DOS is smaller with a V-shaped induced gap, whereas it is U-shaped with a stronger phase modulation in the case of electronic-type carrier density. We attribute this to p-n junctions that form at the graphene/aluminium interface and reduce the contacts' transparency, thus repelling the ABS from the gap edges toward lower energy.

## 6. Derivation of the Supercurrent Spectral Density

In the case of a pair of discrete ABS at energies $\pm E_A$, the DOS reads: $DOS(E, \varphi) = \delta(E + E_A(\varphi)) + \delta(E - E_A(\varphi))$, where $\delta$ is the Dirac distribution and $E_A > 0$ is the Andreev excitation energy that depends on the phase $\varphi$. The energy in the ground state reads $E_{GS}(\varphi) = -E_A(\varphi) = \frac{1}{2}\int_{-\infty}^{\infty} sgn(-E) DOS(E,\varphi) dE$. Following Samuelsson and co-authors[8], the supercurrent spectral density is defined as $J_S(E, \varphi) = \frac{1}{\phi_0}\frac{dE_A}{d\varphi}(-\delta(E + E_A) + \delta(E - E_A))$, where $\phi_0 = \hbar/2e$ is the reduced flux quantum. It quantifies the amount of supercurrent carried by Andreev states at energy $E$ and phase $\varphi$. Therefore, one can see that $J_S(E, \varphi) = -\frac{1}{\phi_0}\int_0^E \frac{\partial DOS}{\partial \varphi}(\epsilon, \varphi) d\epsilon$, the integration boundaries satisfying $J_S(0, \varphi) = 0$. Using this definition, the supercurrent in the ground state reads: $I_S(\varphi) = \frac{1}{2}\int_{-\infty}^{\infty} sgn(-E) J_S(E, \varphi)\, dE = -\frac{1}{\phi_0}\frac{dE_A}{d\varphi}$.

We extend these relations to the case of continuous DOS, which enables one to extract the supercurrent spectral density and the ground state supercurrent from a measured DOS. $J_S$ is a basic quantity naturally expressed in units of $e/h$ and that makes evident the link between Andreev physics and Josephson effect. To get proper units, the tunnelling conductance is converted into a DOS using the proportionality factor $c = 34.72\ meV^{-1}\mu S^{-1}$. We obtained this factor by fitting the V-shaped gate-dependent tunnelling conductance from Fig.S4a to the ideal DOS of graphene in the normal regime $DOS(E_F) = \frac{2\,L\,W}{\pi\,(\hbar v_F)^2}|E_F|$, where the Fermi energy reads $E_F = sgn(V_g)\hbar v_F\sqrt{\pi \frac{c_g|V_g|}{e}}$. In the main text, Fig. 4 shows both the Andreev spectrum in graphene and its corresponding supercurrent spectral density for a gate voltage $V_g = 2.4\,V$.

**Supplementary references**

1. Josephson, B. D. Possible new effects in superconductive tunnelling. *Phys. Lett.* **1,** 251–253 (1962).
2. De Gennes, P. G. Boundary effects in superconductors. *Rev. Mod. Phys.* **36,** 225–237 (1964).
3. Kulik, I. O. Macroscopic quantization and the proximity effect in S-N-S junctions. *Sov. Phys. JETP* **30,** 944–950 (1970).
4. Furusaki, A. & Tsukada, M. Dc Josephson effect and Andreev reflection. *Solid State Commun.* **78,** 299–302 (1991).
5. Beenakker, C. W. J. & van Houten, H. Josephson current through a superconducting quantum point contact shorter than the coherence length. *Phys. Rev. Lett.* **66,** 3056–3059 (1991).
6. Bagwell, P. F. Suppression of the Josephson current through a narrow, mesoscopic, semiconductor channel by a single impurity. *Phys. Rev. B* **46,** 12573–12586 (1992).
7. Wendin, G. & Shumeiko, V. S. Josephson transport in complex mesoscopic structures. *Superlatt. and Microstruct.* **20,** 569–573 (1996).
8. Samuelsson, P., Lantz, J., Shumeiko, V. S. & Wendin, G. Nonequilibrium Josephson current in ballistic multiterminal SNS junctions. *Phys. Rev. B* **62,** 1319–1337 (2000).
9. Le Sueur, H., Joyez, P., Pothier, H., Urbina, C. & Esteve, D. Phase controlled superconducting proximity effect probed by tunneling spectroscopy. *Phys. Rev. Lett.* **100,** 197002 (2008).
10. Pillet, J.-D., Quay, C., Morfin, P., Bena, C. & Levy Yeyati, AJoyez, P. Revealing the electronic structure of a carbon nanotube carrying a supercurrent. *Nat. Phys.* **6,** 965–969 (2010).
11. Chang, W., Manucharyan, V. E., Jespersen, T. S., Nygård, J. & Marcus, C. M. Tunneling spectroscopy of quasiparticle bound states in a spinful josephson junction. *Phys. Rev. Lett.* **110,** 217005 (2013).
12. Bretheau, L., Girit, Ç. Ö., Pothier, H., Esteve, D. & Urbina, C. Exciting Andreev pairs in a superconducting atomic contact. *Nature* **499,** 312–315 (2013).
13. Bretheau, L., Girit, Ç. Ö., Urbina, C., Esteve, D. & Pothier, H. Supercurrent Spectroscopy of Andreev States. *Phys. Rev. X* **3,** 041034 (2013).
14. Castro Neto, A. H., Guinea, F., Peres, N. M. R., Novoselov, K. S. & Geim, A. K. The electronic properties of graphene. *Rev. Mod. Phys.* **81,** 109–162 (2009).
15. Das Sarma, S., Adam, S., Hwang, E. H. & Rossi, E. Electronic transport in two-dimensional graphene. *Rev. Mod. Phys.* **83,** 407–470 (2011).
16. Lindner, N. H., Berg, E., Refael, G. & Stern, A. Fractionalizing majorana fermions: Non-abelian statistics on the edges of abelian quantum hall states. *Phys. Rev. X* **2,** 041002 (2012).
17. Clarke, D. J., Alicea, J. & Shtengel, K. Exotic non-abelian anyons from conventional fractional quantum Hall states. *Nat. Commun.* **4,** 1348 (2013).
18. San-Jose, P., Aguado, R., Guinea, F., Fernandez-Rossier, J. & Lado, J. Majorana Zero Modes in Graphene. *Phys. Rev. X* **5,** 041042 (2015).
19. Calado, V. E. *et al.* Ballistic Josephson junctions in edge-contacted graphene. *Nat. Nanotechnol.* **10,** 761–764 (2015).
20. Ben Shalom, M. *et al.* Quantum oscillations of the critical current and high-field superconducting proximity in ballistic graphene. *Nat. Phys.* **12,** 318–322 (2015).
21. Heersche, H. B., Jarillo-Herrero, P., Oostinga, J. B., Vandersypen, L. M. K. & Morpurgo, A. F. Bipolar supercurrent in graphene. *Nature* **446,** 56–59 (2007).
22. Du, X., Skachko, I. & Andrei, E. Y. Josephson current and multiple Andreev reflections in graphene SNS junctions. *Phys. Rev. B* **77,** 184507 (2008).

23. Girit, Ç. *et al.* Tunable graphene dc superconducting quantum interference device. *Nano Lett.* **9,** 198–199 (2009).
24. Dirks, T. *et al.* Transport Through Andreev Bound States in a Graphene Quantum Dot. *Nat. Phys.* **7,** 386–390 (2011).
25. Komatsu, K., Li, C., Autier-Laurent, S., Bouchiat, H. & Guéron, S. Superconducting proximity effect in long superconductor/graphene/ superconductor junctions: From specular Andreev reflection at zero field to the quantum Hall regime. *Phys. Rev. B* **86,** 115412 (2012).
26. Allen, M. T. *et al.* Spatially resolved edge currents and guided-wave electronic states in graphene. *Nat. Phys.* **12,** 128–133 (2015).
27. Efetov, D. K. *et al.* Specular interband Andreev reflections at van der Waals interfaces between graphene and NbSe2. *Nat. Phys.* **12,** 328–332 (2015).
28. Natterer, F. D. *et al.* Scanning tunneling spectroscopy of proximity superconductivity in epitaxial multilayer graphene. *Phys. Rev. B* **93,** 045406 (2016).
29. Amet, F. *et al.* Supercurrent in the quantum Hall regime. *Science* **352,** 966–969 (2016).
30. Beenakker, C. Specular Andreev Reflection in Graphene. *Phys. Rev. Lett.* **97,** 067007 (2006).
31. Martin, J. *et al.* Observation of electron–hole puddles in graphene using a scanning single-electron transistor. *Nat. Phys.* **4,** 144–148 (2008).
32. Xue, J. M. *et al.* Scanning tunnelling microscopy and spectroscopy of ultra-flat graphene on hexagonal boron nitride. *Nat. Mater.* **10,** 282–285 (2011).
33. van Wees, B. J., Lenssen, K.-M. H. & Harmans, C. J. P. M. Transmission formalism for supercurrent flow in multiprobe superconductor-semiconductor-superconductor devices. *Phys. Rev. B* **44,** 470–473 (1991).
34. Wilhelm, F. K., Schön, G. & Zaikin, A. D. Mesoscopic Superconducting-Normal Metal-Superconducting Transistor. *Phys. Rev. Lett.* **81,** 1682–1685 (1998).
35. Baselmans, J. J. A., Morpurgo, A. F., van Wees, B. J. & Klapwijk, T. M. Reversing the direction of the supercurrent in a controllable Josephson junction. *Nature* **397,** 43–45 (1999).
36. Baselmans, J. J. A., Heikkilä, T. T., van Wees, B. J. & Klapwijk, T. M. Direct observation of the transition from the conventional superconducting state to the pi state in a controllable Josephson junction. *Phys. Rev. Lett.* **89,** 207002 (2002).
37. Wang, L. *et al.* One-dimensional electrical contact to a two-dimensional material. *Science* **342,** 614–7 (2013).
38. Wang, J. I. *et al.* Electronic Transport of Encapsulated Graphene and WSe 2 Devices Fabricated by Pick-up of Prepatterned hBN. *Nano Lett.* **15,** 1898–1903 (2015).
39. Spietz, L., Teufel, J. & Schoelkopf, R. J. A Twisted Pair Cryogenic Filter. *arXiv* 1–12 (2006). Preprint at http://arxiv.org/abs/cond-mat/0601316.
40. Wang, L. *et al.* One-dimensional electrical contact to a two-dimensional material. *Science* **342,** 614–617 (2013).
41. Zomer, P. J., Guimarães, M. H. D., Brant, J. C., Tombros, N. & van Wees, B. J. Fast pick up technique for high quality heterostructures of bilayer graphene and hexagonal boron nitride. *Appl. Phys. Lett.* **105,** 013101 (2014).
42. Gorbachev, R. V *et al.* Hunting for monolayer boron nitride: optical and Raman signatures. *Small* **7,** 465–468 (2011).
43. Amet, F. *et al.* Tunneling spectroscopy of graphene-boron-nitride heterostructures. *Phys. Rev. B* **85,** 073405 (2012).

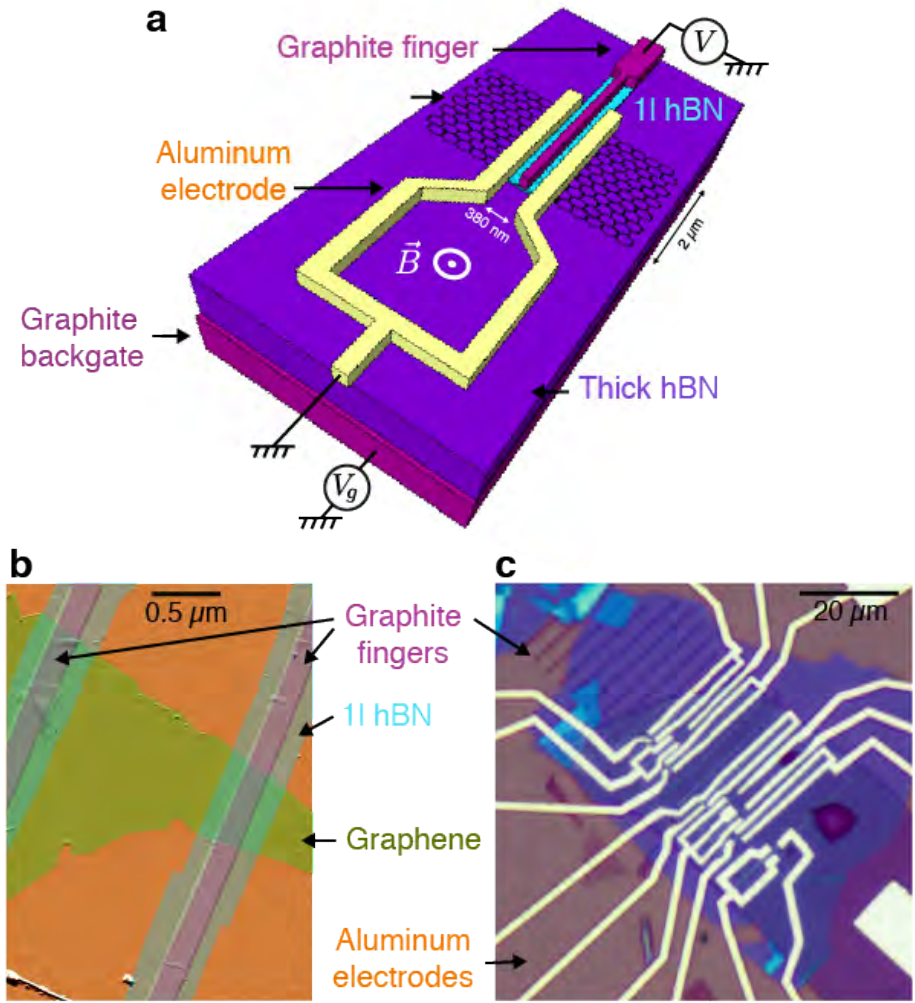

**Figure S1. Van der Waals Heterostructure. a,** Device schematics. **b,** AFM image of the stack. **c,** Optical picture of the device after electron-beam lithography. An encapsulated graphene flake is connected to two superconducting electrodes. On top of it, an insulating barrier of hBN (monolayer) and a graphite electrode enable us to perform tunnelling spectroscopy. The graphite backgate controls electrostatically carrier density in graphene. Magnetic flux $\phi$ threading the superconducting loop imposes a phase $\varphi = \phi/\phi_0$ across graphene.

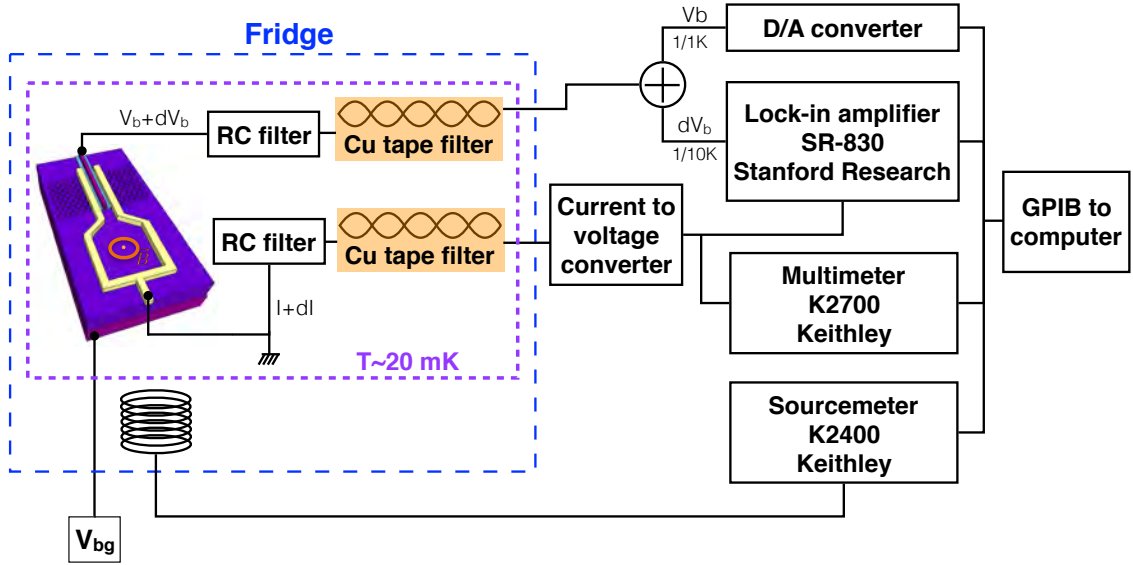

**Figure S2. Measurement Setup.** The sample is anchored to the mixing chamber of a dilution refrigerator at 20 mK. All DC lines are heavily filtered using both discrete RC filters and copper tape filters, which are thermally anchored at 20 mK. *dI/dV* measurements are performed at low frequency (10-100 Hz) using room-temperature amplification and standard lock-in techniques with an excitation voltage of 5-10 µV.

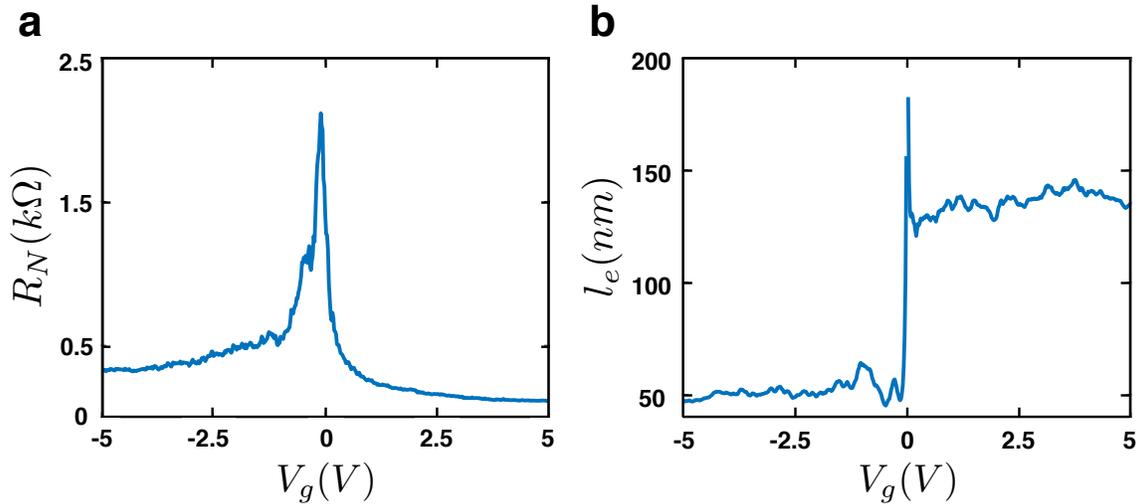

**Figure S3. Scattering Characterization via Transport Measurements. a,** Differential resistance as a function of backgate voltage in the normal regime. This measurement was performed on a similarly fabricated graphene junction between two neighboring SQUID loops. **b,** Lower bound of the mean free path as a function of backgate voltage. It was extracted using the procedure described in the text.

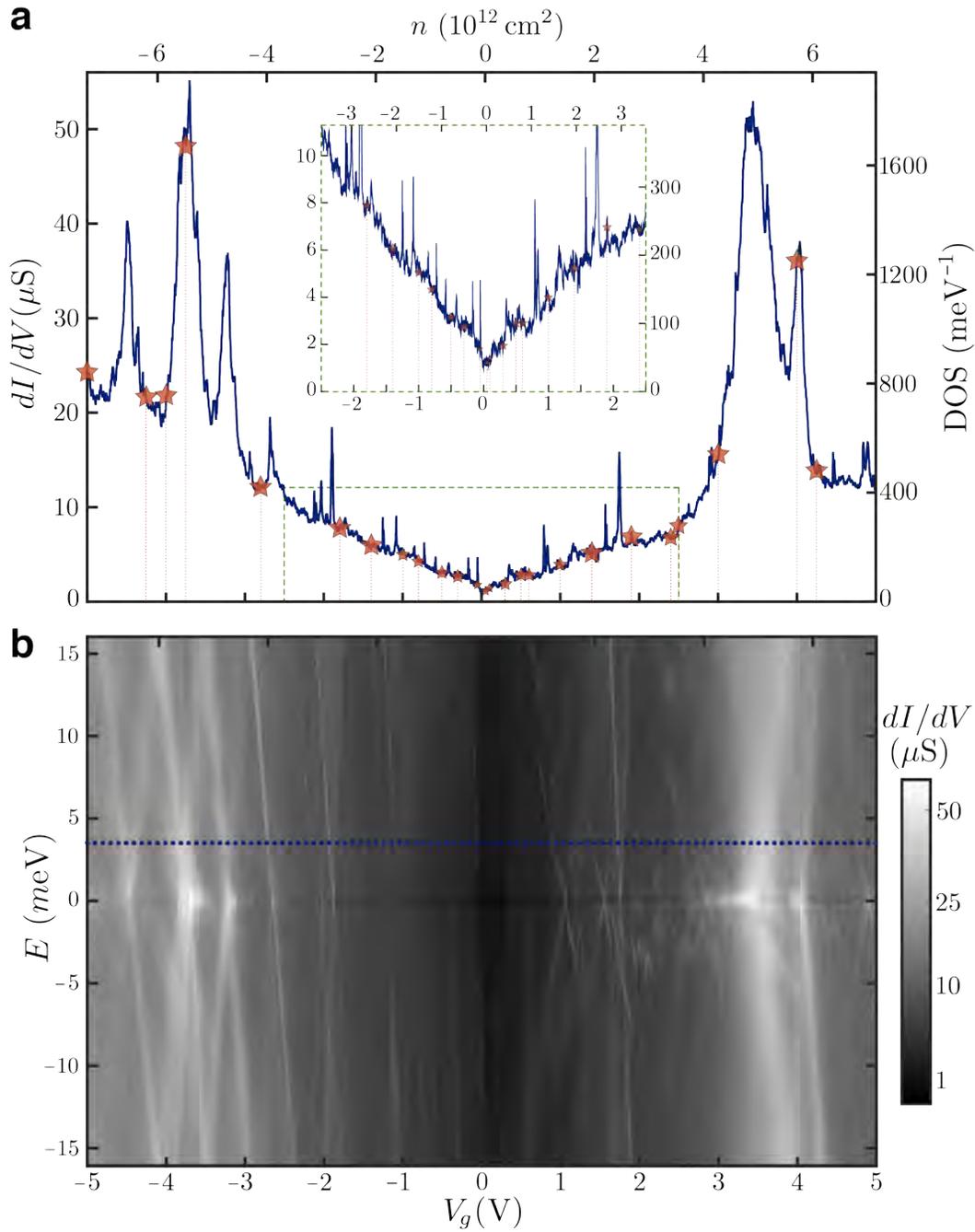

**Figure S4. Graphene's DOS in the Normal State. a, b** Differential conductance, *dI/dV* as a function of gate voltage $V_g$ and energy $E=eV$. On the top x-axis, $V_g$ is converted into graphene's carrier density $n$. On the right y-axis of **(a)**, *dI/dV* is converted into graphene's DOS. **a,** Horizontal line-cut of **(b)** (blue dashed-line) at energy $E = 4\ meV$. Inset: Blow-up around the charge neutrality point. The orange dotted lines show the gate voltage values at which the DOS in the superconducting regime is measured in Fig. S5,6. **(b),** The differential conductance is gray-coded using a logarithmic scale.

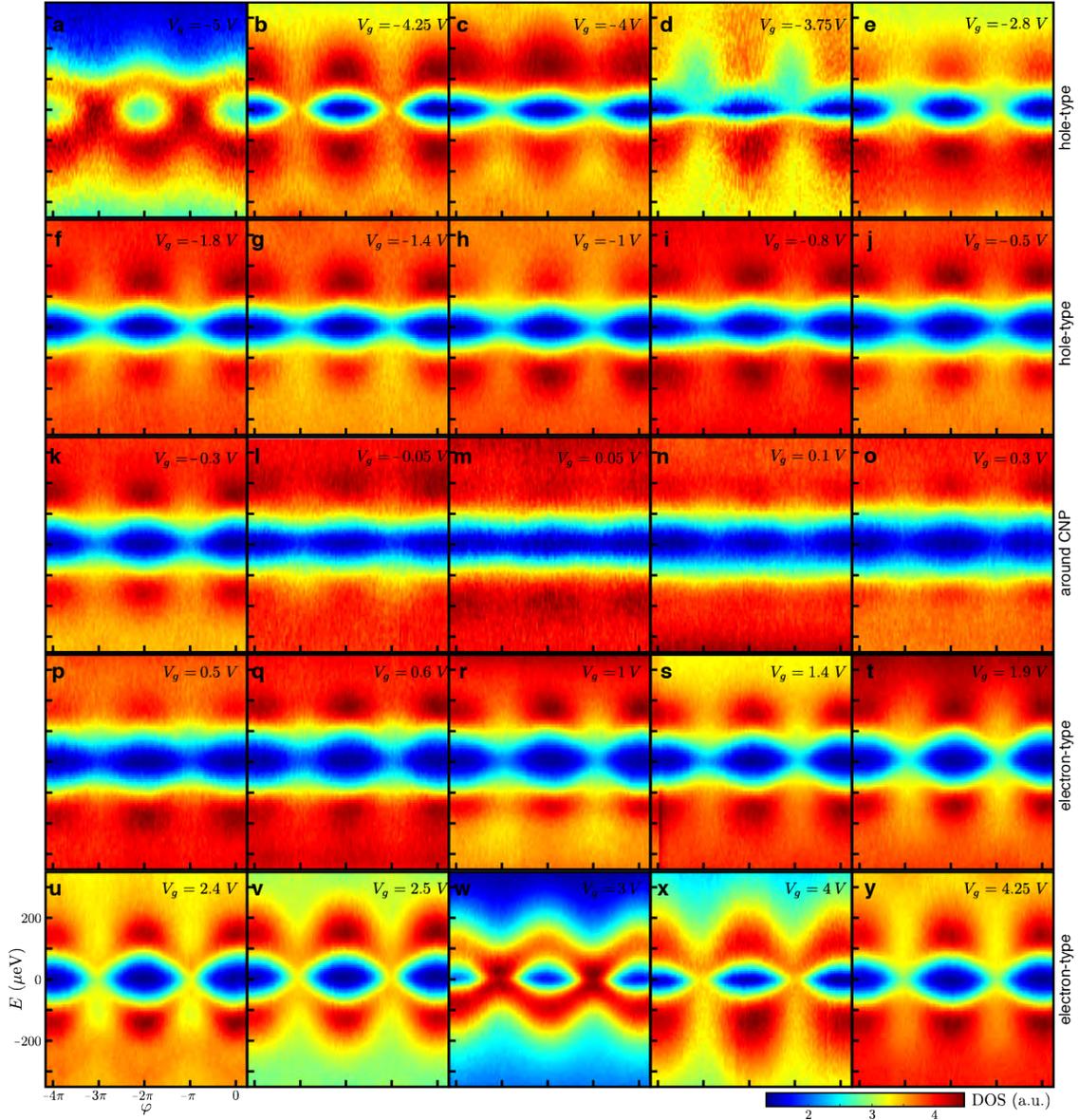

**Figure S5. Gate dependence of the proximitized graphene DOS. a-y**, Colour-coded DOS as a function of both energy $E=eV$ and superconducting phase $\varphi$, for different gate voltages (indicated in each panel). In each panel, the colour-coded DOS is linearly scaled to maximize the contrast (see Fig.S4 for quantitative values).

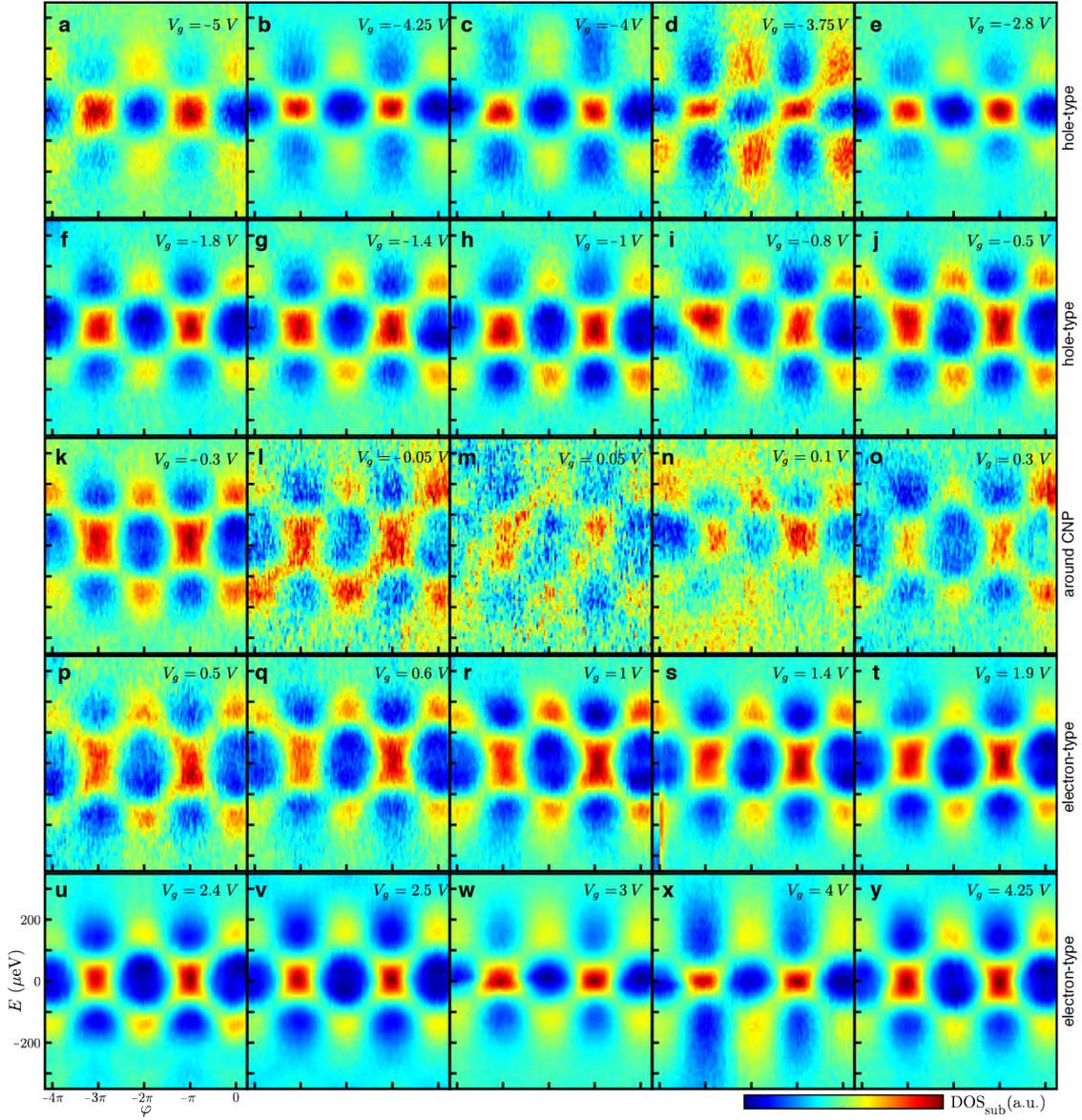

**Figure S6. Gate dependence of subtracted spectra. a-y**, Colour-coded subtracted DOS as a function of both energy $E=eV$ and superconducting phase $\varphi$, for different gate voltages (indicated in each panel). In each panel, the colour-coded subtracted DOS is linearly scaled to maximize the contrast.

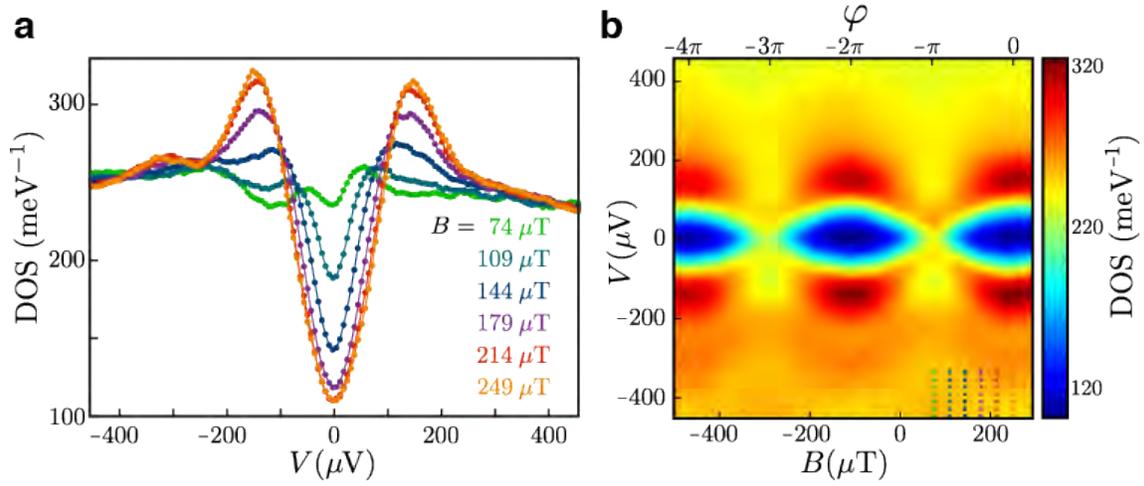

**Figure S7. Phase dependence of graphene's proximitized DOS at $V_g = 2.4\,V$. a, b**, DOS as a function of bias voltage and magnetic field. The induced superconducting gap fully disappears at $\varphi = \pi$ as more ballistic ABS reach zero energy.

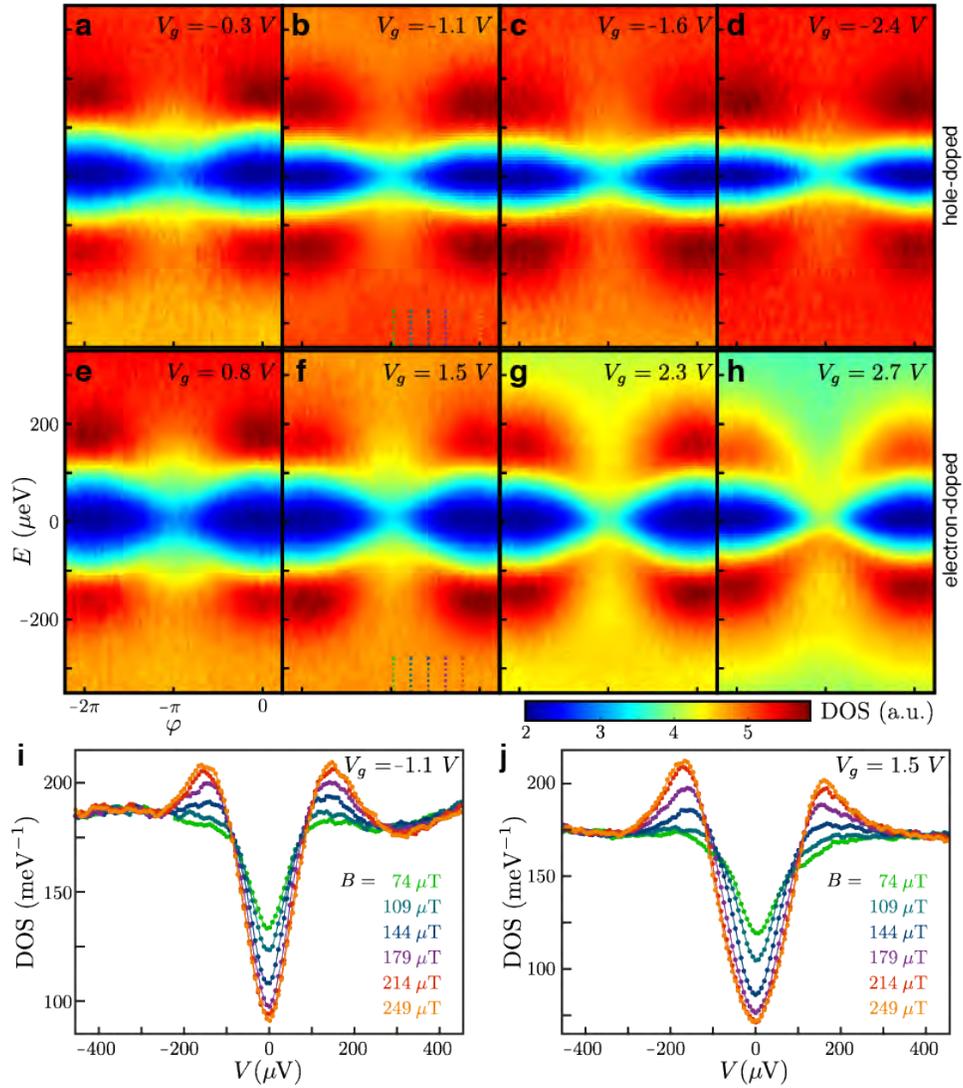

**Figure S8. Comparison of Andreev spectra in the electron and hole-doped regimes.** DOS as a function of both energy $E=eV$ and superconducting phase $\varphi$, for different gate voltages (indicated in each panel) in the hole-doped regime (**a-d, i**) and electron-doped regime (**e-h, j**).